\documentclass[11pt,oneside]{article}
\usepackage{graphicx}
\usepackage[figuresright]{rotating}
\newcommand{\beq}{\begin{equation}}
\newcommand{\eeq}{\end{equation}}

\newcommand{\anu}{\overline {\nu}}

\newcommand{\AmS}{{\protect\the\textfont2
  A\kern-.1667em\lower.5ex\hbox{M}\kern-.125emS}}
\begin{document}
\title{{\bf Neutrino cross sections  and nuclear structure}}
\author{Giampaolo Co', Viviana De Donno, Chiara Maieron\\
\\
\small 
Dipartimento di Fisica, Universit\`a del Salento and\\
\small 
 Istituto Nazionale di Fisica Nucleare sez. di Lecce,\\
\small 
I-73100 Lecce, Italy\\
}
\maketitle
\begin{abstract}
The effects of the theoretical uncertainties in the description of the
neutrino-nucleus cross sections for supernova neutrino energies 
are investigated.
\end{abstract} 
\vskip 0.4cm
The possibility of using neutrinos to make precision astrophysics
measurements is not only related to our capability of detecting them,
but also to the reliability of our theoretical description of the
interaction between neutrinos and electrons and nuclei forming the
detector.  While the neutrino-electron cross section is well described
in terms of perturbation theory, the situation becomes more
complicated when nucleons and nuclei are involved. The internal
structure of these systems is ruled by the strong interaction which is
highly non perturbative in the energetic regime of interest for the
neutrinos emitted from the stars.

We have investigated the consequences of the nuclear structure
uncertainties on the neutrino-nucleus cross sections on the detection
of supernova neutrinos. Computer simulations indicate that the
fluences of the energies transported by these neutrinos, and
antineutrinos, have thermal-like distributions, with average energies
of about 15-30 MeV. The values of the temperature of the electron
neutrinos and antineutrinos fluences are smaller than those
characterizing $\mu$ and $\tau$ neutrinos, since, at the energies
involved, only electron neutrinos, and antineutrinos,
can interact with matter through charge current processes.

We have studied the possibility
of disentangling the temperatures of the $\mu$ and $\tau$ neutrinos
and antineutrinos fluences from that of the electron neutrinos and
antineutrinos. 
%
%
\begin{figure}[t]
\begin{center}
\includegraphics [angle=0.0, scale=0.55]
                 {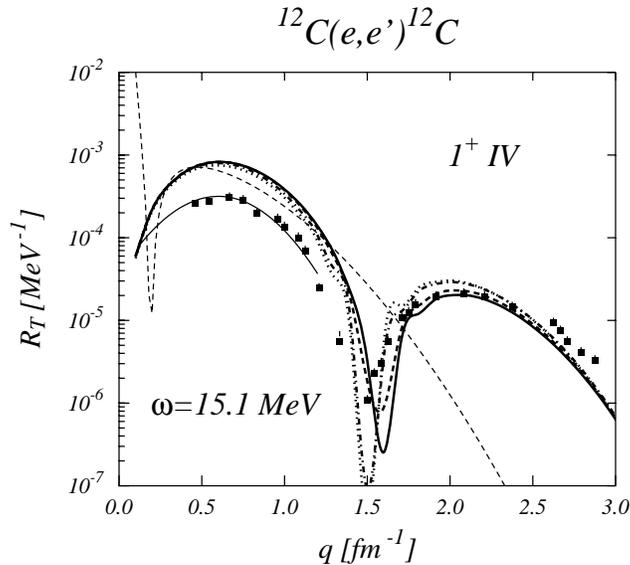} 
\caption{\small Electron scattering form factors for the
  isovector 1$^{+}$ state in $^{12}$C. The different thick lines have
  been obtained by using various effective interactions.  The data are 
  from
  Ref. \cite{hic84}.  The continuous thin line indicates our fit to the
  data. The dashed thin line shows the form factor used in Ref.
  \cite{fuk88}.
  }
\label{fig:ee1+} 
\end{center}
\end{figure}
We have calculated neutrino-nucleus cross sections by using the
effective theory called Random Phase Approximation (RPA) which
describes the excited states of the nucleus as a linear combination of
one-particle one-hole and one-hole one-particle excitations.  The
amplitudes of the linear combination are determined by solving a set
of secular equations depending on the nucleon-nucleon effective
interaction.
We have constructed four effective interactions which reproduce with
the same degree of accuracy the excitation energies of some specific
states and some electromagnetic properties of the double magic nuclei
$^{12}$C, $^{16}$O, $^{40}$Ca, $^{48}$Ca, $^{90}$Zr and $^{208}$Pb. A
detailed description of the interactions and of the procedure used to
construct them is given in \cite{don08t}.  Our study has been done by
making RPA calculations with the four interactions and investigating
the differences produced in various observables. 

In our study we have observed a large sensitivity of the
neutrino-nucleus cross section to the choice the residual interaction,
when the nuclear excitation energy is above the nucleon emission
threshold, as in the giant resonances region. For this reason, here, 
we
address our attention to the excitation of discrete states, and
specifically to the case of $^{12}$C nucleus. We have calculated the
$^{12}$C $(\nu,\nu')$ $^{12}$C cross section for all the multipole
excitations up to angular momentum $J=4$, and we have found that the
main contribution is provided by the excitation of the isovector 1$^+$
state. This state belongs to a triplet of isovector 1$^+$ states. Two
of them are the ground states of the $^{12}$B and $^{12}$N nuclei.
These states can be reached from the $^{12}$C ground state by means of
the reactions $^{12}$C($\anu_e,e^+$)$^{12}$B and
$^{12}$C($\nu_e,e^-$)$^{12}$N which can be identified by measuring the
electron and positron. Also the neutral current reaction exciting the
1$^+$ state can be identified by detecting the emitted photon having 
15.1 MeV energy.
\begin{figure}[t]
\begin{center}
\includegraphics [angle=90.0, scale=0.58]
                 {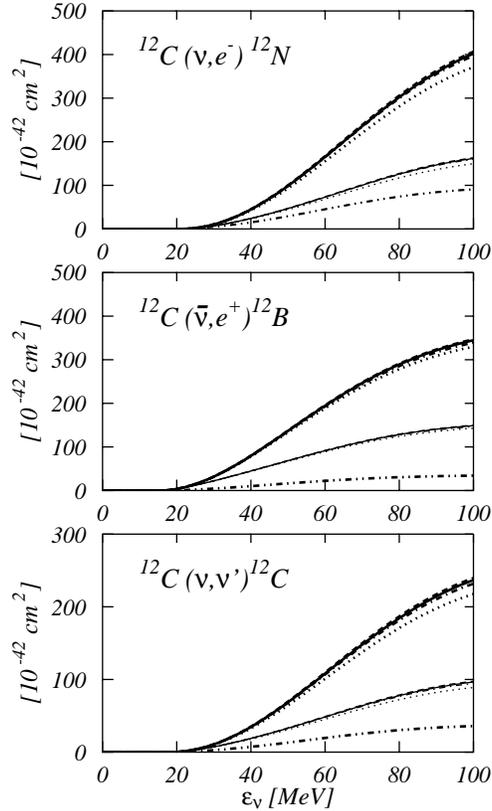}
\caption{\small Neutrino cross sections for reactions exciting 
the isovector 1$^+$ triplet of states in $^{12}$C. 
}
\label{fig:3xsc12} 
\end{center}
\end{figure}

In Fig. \ref{fig:ee1+} we compare the results of our calculations with
the inelastic scattering data of Ref. \cite{hic84} for the isovector
1$^+$ state at 15.1 MeV.  Since the transverse nuclear current matrix
elements of the electron scattering cross sections are also present in
the weak vector current contribution to the neutrino scattering cross
section, we have taken them from experiment, by constructing momentum
dependent quenching functions
\beq
Q(q)=  \frac {<J_f|| T^{exp}_J(q) || J_i>|^2_{e,e'}}
             { |<J_f|| T^V_J(q) || J_i>|^2_{RPA} }
\,\,\,\,,
\label{eq:scal}
\eeq
which we use to rescale the neutrino cross sections.
This procedure has been applied below $q=1$
fm$^{-1}$, i.e. in the kinematic region of interest for supernova
neutrinos. In Fig. \ref{fig:ee1+} we also show the form factor
used in Ref. \cite{fuk88}.

In Fig. \ref{fig:3xsc12} we show the cross sections for the charge
exchange and charge conserving excitation of the 1$^+$ isovector
triplet in $^{12}$C. The higher cross sections are those obtained in
the bare RPA calculations, the set of lower cross section has been
obtained after the rescaling procedure. The cross sections of Ref.
\cite{fuk88} are the lowest ones. We have used these cross sections to
estimate the expected number of events for the various neutrino
interactions detected in the Large Volume Detector located at the Gran
Sasso Laboratory \cite{aga07}. In our estimate we have used the
following values for the supernova parameters: total energy
transported by the neutrinos $E_B \sim 5.0 \,\, 10^{52}$ erg = 8.01
10$^{60}$ MeV, distance from the earth D = 10 kpc, fraction of the
energy carried by a specific type of neutrino, or antineutino,
$f_i$=1/6. This last assumption implies that the energy is equally
divided between the various neutrino types.

The total number of neutral current events $N_{NC}$ is
obtained as the sum of the electron neutrino events calculated for a
fluence with characteristic temperature of 4 MeV, plus the electron
antineutrino having characteristic temperature of 5 MeV, plus the
$\mu$ and $\tau$ neutrinos and antineutrinos  events $N^T_{NC}$ having
all the same temperature $T$. 

\begin{figure}[ht]
\begin{center}
\includegraphics [angle=0.0, scale=0.6]
                 {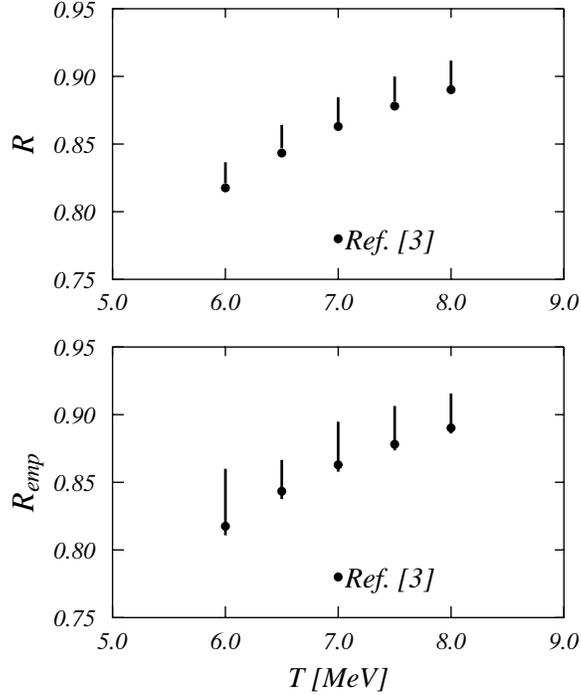}
\caption{\small Upper panel. The ratio $R$ for
some values of the characteristic temperature. Lower panel.
The ratio $R_{emp}$ of Eq. (\ref{eq:rat1}). The points in this case
indicate the results one should obtain. 
}
\label{fig:rat} 
\end{center}
\end{figure}

The results of Fig. \ref{fig:3xsc12}
indicate that there is a noticeable uncertainty on the size of the
cross section, and consequently on the number of expected events.  In
order to reduce this uncertainty we consider the ratio $R =
N^{T}_{NC} / N_{NC}$, shown in the upper panel of
Fig. \ref{fig:rat}, as a function of $T$. 
%
%

The upper and lower values of the bars shown in Fig. \ref{fig:rat}
indicate the largest and smaller value of $R$ calculated with our
cross sections.  The dots represent the values obtained with the cross
sections of Ref. \cite{fuk88}.  The results of the upper panel of Fig.
\ref{fig:rat} show the stability of this new observable 
against the nuclear structure uncertainties.

The experiment is not able to distinguish between neutral current
events produced by electron neutrinos and antineutrinos and those
induced by the other neutrinos and antineutrino types, but it 
has the information on the charge current events which can
be induced only by electron neutrinos, and antineutrinos. 
We simulate an experimental case, by considering the count rates
obtained with the cross sections of Ref. \cite{fuk88} as the detected
counts N$^{emp}$. The ratio to be calculated is 
\begin{eqnarray}
\nonumber
&~& R_{emp} = \frac{1} { N^{emp}_{NC} }  
\Big[ N^{emp}_{NC} \\
\nonumber
&-& N^{emp}_{np} 
\frac{ \displaystyle
  {\int_{e_{th}}^\infty} 
  f(\epsilon)   \sigma_{\nu_e,\nu'_e}(\epsilon) \, d\epsilon 
   }
{\eta_{np} \displaystyle \int_{e_{th}}^\infty  f(\epsilon)
  \sigma_{\nu_e,e^-}(\epsilon)  \, d\epsilon } \\
&-& N^{emp}_{pn}  
  \frac  {\displaystyle \int_{e_{th}}^\infty 
   f(\epsilon)
  \sigma_{\anu_e,\anu'_e}(\epsilon) \, d\epsilon }
   {\displaystyle \eta_{pn}
   \displaystyle \int_{e_{th}}^\infty 
   f(\epsilon)
  \sigma_{\anu_e,e^+}(\epsilon) \, d\epsilon }
\Big] 
\label{eq:rat1}
\,\,\,\,.
\end{eqnarray}
The terms which multiply $N^{emp}_{pn}$ and  $N^{emp}_{np}$ depend on
the knowledge of the neutrino-nucleus cross section, therefore $R_{emp}$
is a model dependent quantity.

 In the lower panel of Fig. \ref{fig:rat} 
the spreading of the $R_{emp}$ calculated with the various
interactions and with the rescaled cross sections are indicated by the
various bars. The correct values of $R_{emp}$, obtained by inserting
the cross sections of Ref. \cite{fuk88} in Eq. \ref{eq:rat1} are
indicated by the dots. 
In conclusion, the uncertainty on the theoretical neutrino-nucleus
cross sections in the giant resonance region is too large to make
reliable predictions for astrophysical purposes. The discrete
excitations are much better described, even though there are problems
on the overall size of the cross sections. The use of ratio of
observables reduces the effects of the nuclear structure
uncertainties.
%
%
%

%
\end{document}